# Automated Dynamic Offset Applied to Cell Association

Majed Haddad*, Habib Sidi†, Piotr Wiecek‡ and Eitan Altman*
*INRIA Sophia-Antipolis, Sophia-Antipolis, France
†CERI/LIA, University of Avignon, Avignon, France
‡Institute of Mathematics and Computer Science, Wroclaw University of Technology, Poland

*Abstract*—In this paper, we develop a hierarchical Bayesian game framework for automated dynamic offset selection. Users compete to maximize their throughput by picking the best locally serving radio access network (RAN) with respect to their own measurement, their demand and a partial statistical channel state information (CSI) of other users. In particular, we investigate the properties of a Stackelberg game, in which the base station is a player on its own. We derive analytically the utilities related to the channel quality perceived by users to obtain the equilibria. We study the Price of Anarchy (PoA) of such system, where the PoA is the ratio of the social welfare attained when a network planner chooses policies to maximize social welfare versus the social welfare attained in Nash/Stackeleberg equilibrium when users choose their policies strategically. We show by means of a Stackelberg formulation, how the operator, by sending appropriate information about the state of the channel, can configure a dynamic offset that optimizes its global utility while users maximize their individual utilities. The proposed hierarchical decision approach for wireless networks can reach a good trade-off between the global network performance at the equilibrium and the requested amount of signaling. Typically, it is shown that when the network goal is orthogonal to user's goal, this can lead the users to a misleading association problem.

*Index Terms*—WLAN, 3G, association problem, misleading information, channel state information, game theory, Bayes-Nash equilibrium, Bayes-Stackelberg equilibrium, Price of Anarchy.

## I. INTRODUCTION

Efficient design of wireless networks calls for end users implementing radio resource management (RRM), which requires knowledge of the mutual channel state information in order to limit the influence of interference impairments on the decision making. However, full CSI assumption is not always practical because communicating channel gains between different users in a time varying channel within the channel coherence time may lead to large overhead. In this case, it is more appropriate to consider each channel coherence time as a one-stage game where players are only aware of their own channel gains and their opponent's channel statistics (which vary slowly compared to the channel gains and, therefore, can be communicated [1]). The interaction between the players may be repeated but with a different and independent channel realization each time and therefore is not a repeated game. This motivates the use of games with incomplete information, also known as Bayesian games [2], [3] which have been incorporated into wireless communications for problems such as power control [4] and spectrum management in the interference channel [5]. In [4], a distributed uplink power control in a multiple access (MAC) fading channel was studied and shown to have a unique Nash equilibrium (NE) point. With the same incomplete information, it was shown [5] that in a symmetric interference channel with a one-time interaction, there exists a unique symmetric strategy profile which is a NE point. This result however is limited to scenarios where all users statistically experience identical channel conditions (due to the symmetry assumption) and does not apply to interactions between weak and strong users.

In this paper, we present an alternative approach for configuring a dynamic offset by introducing a certain degree of hierarchy between the users and the base station. More specifically, we propose a Stackelberg formulation of the association problem when a partial channel state information is assumed at the transmitter. By Stackelberg we mean distributed decision making assisted by the network, where the wireless users aim at maximizing their own utility, guided by aggregated information broadcasted by the network about the CSI of each user. We first show how to derive the utilities of users that are related to their respective channel quality under the different association policies. We then derive the policy that corresponds to the Stackelberg equilibrium and compare it to the centralized and the non-cooperative model. Technically, our approach not only aims at improving the network equilibrium efficiency but has also two nice features: (i) It allows the network to guide users to a desired equilibrium that optimizes its own utility if it chooses the adequate information to send, (ii) Only the individual user demand and a partial statistical CSI of other users is needed at each transmitter. Our approach contributes to designing networks where intelligence is split between the base station (BS) and mobile stations (MSs) in order to find a desired trade-off between the global network performance reached at the equilibrium and the amount of signaling needed to make it work. Note that the Stackelberg formulation arises naturally in some contexts of practical interest. For example, hierarchy is naturally present in contexts where there are primary (licensed) users and secondary (unlicensed) users who can sense their environment because there are equipped with a radio [6]. It is also natural if the users have access to the medium in an asynchronous manner.

Moreover, this game has an unusual information: it is *partial* and ***misleading***. Misleading - because, although the channel state indeed can give information on the transmission rate, it is known that the actual throughput of a given user is a function of not only his channel state but also of that of the other connected users [7]. The throughput is known to be lower bounded by the harmonic mean of the rates available to each user [8]. Thus the nodes transmitting at higher rate is degraded below the level of lower bit rate. The real utility of a given user is the throughput he would get and the user may not be aware that it is possible that an access point with a better channel may have a lower throughput because more terminals are connected to it.

From the system design perspective, the given model is very useful in practice. Recently, papers like [9] raises the cell reselection process problem in HetNets. Clearly, the association based on highest signal strength is inadequate to address this challenge. Moreover, in the 3GPP RRC standard [10] in Section 5.1.5 it is clearly said that the E-UTRAN can configure a list of cell specific offsets and a list of *blacklisted* cells. Typically, Release 8 UEs should apply the ranking based on radio link quality (with offsets) unless operator indicates support for priority-based reselection. This suggests that the offset can be configured for each and every cell depending on the priority set by the operator (e.g., mobility, requested throughput, amount of signalling).

## II. SYSTEM MODEL

Consider a heterogeneous wireless system consisting of a single MAN (Metropolitan Area Network, e.g. 3G) cell and an overlapping wireless LAN (Local Area Networks, e.g. WiFi) hot-spot. Each user entering in the system will decide *individually* to which of the available systems it is best to connect according to its radio condition, its demand and the statistical information about other users. Their policies (or strategies) are then based on this (incomplete) information. The association problem is then generalized to allow the BS to control the users' behavior by broadcasting appropriate information, expected to maximize its utility while individual users maximize their own utility.

$\sigma^2$ is the noise variance. $b_i$ is the demand of user $i$ ($b_i = 1$ when there exists a demand, and $0$ otherwise) and $a_i$ his action defined by the user decision to connect to a certain radio access technology (RAT). $a_i = 1$ when the user chooses 3G, and $0$ when the user chooses WiFi. $h_i$ is the downlink channel power gain between the 3G BS and the end-terminal. The channel gains could be either independent or correlated over the $n$ users. We assume that all channels are independent and undergo Rayleigh fading. By the transformation theorem for single random variable, the channel power gain $h_i$ has an exponential distribution with mean $\lambda_i$ [11]. We will see later how $\lambda_i$ is related to different parameters adopted throughout the paper. We assume that the user state is defined by the pair $(h_i, b_i)$. The network is fully characterized by the user state. However, when distributing the JRRM decisions, this complete information is not available to the users. The BS and the AP broadcasts to its terminals an aggregated information indicating a measurement of the communication quality of the wireless channel (excellent, fair, poor...). This can be done through an offset which we will also call in the sequel interchangeably as the Channel Quality Indicator (CQI). The CQI can be a value (or values) representing a measure of channel quality for a given channel. Typically, a high value CQI is indicative of a channel with high quality and vice versa. More formally, assume that the knowledge of each user about his own state is limited to the pair $(s_i, b_i)$, where $s_i = 1\!\!\mathrm{I}_{\{h_i > \Psi_i\}}$, with $\Psi$ – a fixed threshold which represents the dynamic offset parameter set by the 3G BS and $1\!\!\mathrm{I}_C$ is the indicator function equal to 1 if condition $C$ is satisfied and to 0 otherwise. We will call $\Psi_i$ the "*CQI threshold*" of user $i$. Thus, a user only knows whether he wants to transmit and whether the channel is in a good ($s_i = 1$) or in a bad ($s_i = 0$) condition given the CQI threshold. In addition any player has the information about the probability distribution of his own state $(s_i, b_i)$ and that of his opponent $(s_j, b_j)$. These are given by $\alpha_i$ – the probability to have $\{h_i > \Psi_i\}$, and $\beta_i$ – the probability that $b_i = 1$. Let us denote by $\mathbf{P} = [\mathbf{P_1}, \ldots, \mathbf{P_n}]^T$ the $(n \times 2)$ policy profile matrix, whose element $\mathbf{P_i}$ represents the action vector taken by the mobile $i$ in low and high channel states for $i = 1, \ldots, n$.

In the next sections, we provide a thorough analysis of the existence and characterization of the Bayes equilibria for both non-cooperative and Stackelberg scenarios. We first focus on the two-user case in order to gain insights into how to design decision problem in radio environments. Then, we generalize our approach to the multi-user case.

### A. WiFi Throughput

The measurement of average throughput of a node in a wireless LAN is done by the time it takes to transfer the files between the WiFi access point (AP) and the wireless clients. Typically, one would transfer a file from a wired server to a wireless client by means of an AP bridging wired and wireless networks. The throughput depends on the bit rate at which the wireless mobile communicates to its AP. On the other hand, as already mentioned, if there is at least one host with a lower rate, a WLAN network presents a performance anomaly in the sense that the throughput of all the hosts transmitting at higher rate is degraded below the level of the lower rate [7], [8], [12]. We can accordingly consider that the throughput of a WiFi connection is equal to a constant, say $v$, regardless of differences in users' channel data rate.

### B. 3G Throughput

As opposed to WiFi, the 3G technology uses CDMA multiplexing. Hence, each user receives a certain number of codes which are converted into a certain amount of throughput depending on the chosen modulation and coding scheme, which greatly depends on the link quality at the receiver side. This can vary greatly depending on the link conditions due to interference and noise impairments. We then model the utility experienced by a user that is connected to 3G by the capacity of Shannon [13]. Assuming that there is no

interference between 3G and the WiFi network, the throughput of a user connected to 3G system is given by:

$$Thp_i^C = \log\left(1 + \frac{p\,h_i\,a_i\,b_i}{\sigma^2 + p\,h_j\,a_j\,b_j}\right); \qquad j \neq i \quad (1)$$

where index $C$ stands for 3G cellular network.

As can be seen, the throughput obtained by a user in a system depends on both his own decisions and the decisions taken by the other users. Given $\lambda_i$ and $\Psi_i$, we can compute that the distribution of $h_i$ is $\text{Exp}(\lambda_i)$ with

$$\alpha_i = \exp(-\lambda_i \Psi_i) \quad (2)$$

Given the information that a player has, there are four possible policies of a player $i$ with $b_i = 1$ (we do not consider state $b_i = 0$, when there is no transmission of any type):

| $h_i < \Psi_i$ | W | W | C | C |
|---|---|---|---|---|
| $h_i > \Psi_i$ | W | C | W | C |

where index $W$ stands for WiFi network. Let us not consider the policy $(C, W)$, which is irrational, as the throughput of a player using 3G when $\{h_i > \Psi_i\}$ is certainly higher than that when $\{h_i < \Psi_i\}$. We then have a game with partial CSI with two states and a $(3 \times 3)$ matrix in every state.

For the ease of comprehension, we will begin by considering the two-user case and then generalize the results to the multi-user case later in Section IV.

### III. THE TWO-USER CASE

User $i$'s utility in state $s = 0, 1$ is given by

$$u_i(s, \mathbf{P}) = \begin{cases} v; & \text{if user } i \text{ chooses } W \text{ at state } s, \\ C^i_{\mathbf{P_j}}(s); & \text{if user } i \text{ chooses } C \text{ at state } s \end{cases} \quad (3)$$

The functions $C^i_k$, describing the utility of player $i$ using 3G when his opponent applies policy $k$, are defined as follows

$$C^i_k(1) = \mathbb{E}[c^i_k(h_i) | h_i > \Psi_i] = \frac{1}{\alpha_i} \int_{\Psi_i}^{\infty} c^i_k(h_i) \lambda_i e^{-\lambda_i h_i}\, dh_i,$$

$$C^i_k(0) = \mathbb{E}[c^i_k(h_i) | h_i < \Psi_i] = \frac{1}{1 - \alpha_i} \int_0^{\Psi_i} c^i_k(h_i) \lambda_i e^{-\lambda_i h_i}\, dh_i,$$

with $k = WW, WC, CC$.

$c^i_k(h_i)$ above is the utility of player $i$ using $C$ when channel gain is $h_i$ against policy $k$ of player $j$. These utilities are defined as follows:

$$c^i_{CC}(h_i) = \beta_j \int_0^\infty \log\left(1 + \frac{ph_i}{\sigma^2 + ph_j}\right)\lambda_j e^{-\lambda_j h_j}\,dh_j + (1-\beta_j)\int_0^\infty \log\left(1 + \frac{ph_i}{\sigma^2}\right)\lambda_j e^{-\lambda_j h_j}\,dh_j \quad (4)$$

Next:

$$c^i_{WC}(h_i) = \beta_j \int_{\Psi_j}^\infty \log\left(1 + \frac{ph_i}{\sigma^2 + ph_j}\right)\lambda_j e^{-\lambda_j h_j}\,dh_j + (1-\beta_j)\int_{\Psi_j}^\infty \log\left(1 + \frac{ph_i}{\sigma^2}\right)\lambda_j e^{-\lambda_j h_j}\,dh_j + \int_0^{\Psi_j} \log\left(1 + \frac{ph_i}{\sigma^2}\right)\lambda_j e^{-\lambda_j h_j}\,dh_j \quad (5)$$

Finally:

$$c^i_{WW}(h_i) = \log\left(1 + \frac{ph_i}{\sigma^2}\right) \quad (6)$$

*A. The non-cooperative equilibrium*

Game theory has accentuated the importance of randomized games or mixed games. However, such a game does not find a significant role in most communication modems and source coding codecs since equilibria where each user randomly picks a decision at each time epoch are unfortunately not interesting in such a case, as they amount to perpetual handover between networks. In what follows, we will make use of the users' utilities obtained above to derive the pure association strategies.

**Definition 1** (Bayes-Nash equilibrium). *A strategy profile $\mathbf{P_i}^{BNE}$, $\forall i = 1, 2$ corresponds to a Bayes-Nash equilibrium (BNE) if, for all users, any unilateral switching to a different strategy cannot improve user's payoff at any state. Mathematically, this can be expressed by the following inequality, given the statistical information about the other user $\forall \mathbf{Q_i} \neq \mathbf{P_i}^{BNE}$*

$$u_i(s_i, (\mathbf{P_i}^{BNE}, \boldsymbol{P}^{BNE}_{-i})) \geq u_i(s_i, (\mathbf{Q_i}, \boldsymbol{P}^{BNE}_{-i})); \qquad \text{for } s_i = 0, 1$$

**Proposition 1.** *The game considered in the paper always has a pure-strategy Bayes-Nash equilibrium. Moreover*

(a) *$(CC, CC)$ is an equilibrium iff $C^i_{CC}(0) \geq v$ for $i = 1, 2$.*
(b) *$(CC, WC)$ is an equilibrium iff $C^1_{WC}(0) \geq v$ and $C^2_{CC}(0) \leq v \leq C^2_{CC}(1)$.*
(c) *$(CC, WW)$ is an equilibrium iff $C^1_{WW}(0) \geq v$ and $C^2_{CC}(1) \leq v$.*
(d) *$(WC, CC)$ is an equilibrium iff $C^1_{CC}(0) \leq v \leq C^1_{CC}(1)$ and $C^2_{WC}(0) \geq v$.*
(e) *$(WC, WC)$ is an equilibrium iff $C^i_{WC}(0) \leq v \leq C^i_{WC}(1)$ for $i = 1, 2$.*
(f) *$(WC, WW)$ is an equilibrium iff $C^1_{WW}(0) \leq v \leq C^1_{WW}(1)$ and $C^2_{WC}(1) \leq v$.*
(g) *$(WW, CC)$ is an equilibrium iff $C^1_{CC}(1) \leq v$ and $C^2_{WW}(0) \geq v$.*
(h) *$(WW, WC)$ is an equilibrium iff $C^1_{WC}(1) \leq v$ and $C^2_{WW}(0) \leq v \leq C^2_{WW}(1)$.*
(i) *$(WW, WW)$ is an equilibrium iff $C^i_{WW}(1) \leq v$ for $i = 1, 2$.*

*Proof:* The statements (a)–(i) are direct consequences of the definition of Bayes-Nash equilibrium and the form of payoff matrices. Next, it is immediate to see that the definitions of $C^i_k(s)$ imply the following inequalities:

$$C^i_{CC}(s) < C^i_{WC}(s) < C^i_{WW}(s)$$

for $i=1,2$ and $s=0,1$. Now, using these inequalities, it is tedious but straightforward to show that always at least one of the conditions (a)–(i) is satisfied. ∎

The next proposition gives us some information on how the Nash-Bayes equilibria depend on the chosen values of the CQI thresholds $\Psi_i$. Form a simple analysis, we can come up with the following informations on how the Nash-Bayes equilibria depend on the chosen values of the CQI thresholds $\Psi_i$.

**Proposition 2.** *If $\Psi_1$ and $\Psi_2$ are small enough none of the players uses $CC$ in equilibrium. If they are large enough, none of the players uses policy $WW$ in equilibrium. Moreover, for all the values of the parameters of the model one of the two possibilities is true:*

*(a) For $\Psi_1$ and $\Psi_2$ small enough at least one of the players uses policy $WW$ in equilibrium,*
*(b) For $\Psi_1$ and $\Psi_2$ large enough at least one of the players uses policy $CC$ in equilibrium.*

More discussion on this result is given in [14].

*Proof:* Define for $i=1,2$ and $k=CC,WC,WW$

$$C_k^i(\infty) = \int_0^\infty c_k^i(h_i)\lambda_i e^{-\lambda_i h_i}\,dh_i \qquad (7)$$

Note that when $\Psi_1 \to 0$ and $\Psi_2 \to 0$, $C_k^i(0)(\Psi_1,\Psi_2) \to 0$ and $C_k^i(1)(\Psi_1,\Psi_2) \to C_k^i(\infty)$ for $i=1,2$, $k=CC,WC,WW$. Analogously, when $\Psi_1 \to \infty$ and $\Psi_2 \to \infty$, $C_k^i(0)(\Psi_1,\Psi_2) \to C_k^i(\infty)$ and [1] $C_k^i(1)(\Psi_1,\Psi_2) \to +\infty$. Thus for $\Psi_1$ and $\Psi_2$ small enough, $C_k^i(0)(\Psi_1,\Psi_2) < v$ for all the values of $i$ and $k$, which by Proposition 1 implies that no player uses policy $CC$ in equilibrium. Analogously for $\Psi_1$, $\Psi_2$ big enough, $C_k^i(0)(\Psi_1,\Psi_2) > v$ for all the values of $i$ and $k$, and thus no player uses $WW$ in equilibrium then.

Now note that by Proposition 1, one of the players uses $CC$ in equilibrium iff

$$C_{WC}^1(0) \geq v \quad \text{or} \quad C_{WC}^2(0) \geq v.$$

Thus if we take $\Psi_1$, $\Psi_2$ large enough, we can pass to the limit:

$$C_{WC}^1(\infty) \geq v \quad \text{or} \quad C_{WC}^2(\infty) \geq v. \qquad (8)$$

Analogously, one of the players uses $WW$ in equilibrium iff

$$C_{WC}^1(1) \leq v \quad \text{or} \quad C_{WC}^2(1) \leq v.$$

Passing to the limit when $\Psi_1$, $\Psi_2$ approach 0,

$$C_{WC}^1(\infty) \leq v \quad \text{or} \quad C_{WC}^2(\infty) \leq v. \qquad (9)$$

[1] To prove that $C_k^i(0)(\Psi_1,\Psi_2) \to 0$ and $C_k^i(1)(\Psi_1,\Psi_2) \to +\infty$ it is enough to notice that the conditional expected values (III) and (III) are bounded from above and from below respectively by the biggest value of $c_k^i$ on the set $\{h_i : h_i < \Psi_i\}$ and the smallest one on the set $\{h_i : h_i > \Psi_i\}$. Note however that the former is bounded from above by $\log\left(1+\frac{ph_i}{\sigma^2}\right)$ which converges to 0 as $h_i \to 0$. Then the latter one is bounded from below by $(1-\beta_j)\int_0^\infty \log\left(1+\frac{ph_i}{\sigma^2}\right)\lambda_j e^{-\lambda_j h_j}\,dh_j = (1-\beta_j)\log\left(1+\frac{ph_i}{\sigma^2}\right)$, which clearly goes to infinity as $h_i \to \infty$.

However (8) and (9) cover all the values of $v$, ending the proof. ∎

Roughly speaking, this means that for higher values of the CQI thresholds $\Psi_i$s the players are more likely to use 3G rather than WiFi and conversely, for low values of the CQI thresholds $\Psi_i$s the players are more likely to use WiFi rather than 3G. Interestingly, this result also suggests that, rather than increasing the offered throughput $v$, the operator could control the equilibrium of its wireless users to maximize its own revenue by broadcasting appropriate CQI thresholds. [2]. This can lead the network to minimize its overall cost and users to a misleading association problem. Next, we address this problem by introducing a hierarchical model.

### B. The hierarchical equilibrium

In this section, we propose a methodology that transforms the above non-cooperative game into a Stackelberg game. Concretely, the network may guide users to an equilibrium that optimizes its own utility if it chooses the adequate information to send. We first study the policy that maximizes the utility of the network, which is defined as the expected number of active users connected to 3G network, which for two users can be defined as follows:

$$\begin{aligned}
U_{BS}(\mathbf{P},\Psi_1,\Psi_2) &= \beta_1(\mathbb{1}_{\{\mathbf{P_1}=CC\}} + \alpha_1\mathbb{1}_{\{\mathbf{P_1}=WC\}}) + \\
&\quad \beta_2(\mathbb{1}_{\{\mathbf{P_2}=CC\}} + \alpha_2\mathbb{1}_{\{\mathbf{P_2}=WC\}}) \\
&= \beta_1(\mathbb{1}_{\{\mathbf{P_1}=CC\}} + e^{-\lambda_1\Psi_1}\mathbb{1}_{\{\mathbf{P_1}=WC\}}) + \\
&\quad \beta_2(\mathbb{1}_{\{\mathbf{P_2}=CC\}} + e^{-\lambda_2\Psi_2}\mathbb{1}_{\{\mathbf{P_2}=WC\}})
\end{aligned}$$

Nevertheless, as it is not realistic to consider that the users will seek the global optimum, we show how to find the policy that corresponds to the Bayes-Stackelberg equilibrium where the BS tries to maximize its own utility $U_{BS}$ just by choosing the CQI thresholds, knowing that users will try to maximize their individual utilities.

**Definition 2** (**Bayes-Stackelberg equilibrium**). *By denoting $(\Psi_1^{BSE},\Psi_2^{BSE})$ the strategy profile of the BS at a Bayes-Stackelberg equilibrium (BSE), this definition translates mathematically as*

$$(\Psi_1^{BSE},\Psi_2^{BSE}) \subset \arg\max_{\Psi_1,\Psi_2} U_{BS}(\mathbf{P}^{BNE}(\Psi_1,\Psi_2),\Psi_1,\Psi_2), \qquad (10)$$

*where $\mathbf{P}^{BNE}(\Psi_1,\Psi_2)$ is any Bayes-Nash equilibrium in the game of the previous section with CQI thresholds equal to $\Psi_1$, $\Psi_2$.*

---

[2] Of course there are some limits of the possible influence of the operator. As it is clearly seen from the proof given above, the situation when users always choose WiFi for small values of the given CQI thresholds and 3G for the big ones is not possible – the possible situations are the following: when the value of $\Psi_i$ is small, the player chooses WiFi below this threshold (and so he does it with a low probability) and 3G above this threshold. When $\Psi_i$ gets bigger it ceases to provide any information to the player and thus he chooses 3G both below and above $\Psi_i$, which means that he still chooses 3G with a higher probability. The second possible situation is that for a low value of $\Psi_i$ player $i$ chooses WiFi for both his states, while for a high value of $\Psi_i$ he distinguishes between these states by choosing either 3G or WiFi, but again for any value of the CQI threshold the probability that WiFi is used is smaller than that of using 3G.

We next exemplify our general analysis by investigating the possibility of considering three scenarios for the choice of $\Psi_1$ and $\Psi_2$:

1) *Centralized model* – the base station chooses both $\Psi_i$s and the policies for the players, aiming to maximize the expected number of players using 3G in the second stage. Formally, the centralized strategy is the one satisfying

$$(\Psi_1{}^C, \Psi_2{}^C, \mathbf{P}^C) \subset \arg\max_{\Psi_1, \Psi_2, \mathbf{P}} U_{BS}(\mathbf{P}, \Psi_1, \Psi_2),$$

2) *Stackelberg model* – there are two stages: at the first one the base station chooses both $\Psi_i$s given the information about the distributions of $(h_i, b_i)$ aiming to maximize the expected number of players using 3G at the second stage, when players play the game from the last section. The proposed approach can be seen as intermediate scheme between the centralized model and the fully non-cooperative model,

3) *Fully non-cooperative model* – the game has two stages: at the first one, players choose their $\Psi_i$s given the information they have about the distributions of $(h_i, b_i)$ aiming to maximize their expected throughput at the second stage; at the second stage they choose a policy depending on actual $(s_i, b_i)$ as in the model of the last section. Formally, a fully non-cooperative strategy is any one satisfying

$$\Psi_i^{NC} \subset \arg\max_{\Psi_i} \mathbb{E}[u_i(s_i, \mathbf{P}^{BNE}(\Psi_i, \Psi_j^{NC}))]; \text{ for } i = 1, 2$$

with $\mathbf{P}^{BNE}(\Psi_i, \Psi_j^{NC})$ being any Bayes-Nash equilibrium in the game of the previous section.

Below, we analyze the behavior of the base station and the players at the equilibria of each of these models.

**Proposition 3.**

1) *In the centralized model, the base station chooses any $\Psi_1$ and $\Psi_2$, and $CC$ policies for both users.*

2) *In the Stackelberg model, when $C_{CC}^i(\infty) > v$ for $i = 1, 2$ then the base station chooses any $\Psi_1 > \Psi_1^{**}$ and $\Psi_2 > \Psi_2^{**}$ with $\Psi_i^{**}$ satisfying[3] $C_{CC}^1(0)(\Psi_1^{**}) = v$ and $C_{CC}^2(0)(\Psi_2^{**}) = v$ and then users both play $CC$. When $C_{CC}^i(\infty) \leq v$ for $i = 1$ or $i = 2$, then the base station chooses $\Psi_i^{***}$, for $i = 1, 2$ maximizing[4] either*

$$\beta_1 e^{-\lambda_1 \Psi_1} + \beta_2 e^{-\lambda_2 \Psi_2}$$

*subject to $C_{WC}^i(1)(\Psi_1, \Psi_2) \geq v \geq C_{WC}^i(0)(\Psi_1, \Psi_2)$, $i = 1, 2$ or maximizing*

$$\beta_i + \beta_j e^{-\lambda_j \Psi_j}$$

*subject to $C_{WC}^i(0)(\Psi_1, \Psi_2) \geq v$ and $C_{CC}^j(1)(\Psi_1, \Psi_2) \geq v$. In the first case, both players choose $WC$ in the second stage. In the second case, user $j$ chooses $WC$ and user $i$ chooses $CC$.*

---
[3]Here and in the sequel $C_k^i(s)(\Psi_1, \Psi_2)$ denotes the respective $C_k^i(s)$ when the values of $\Psi_i$s have the given value.

[4]Of course the one of the two with the higher objective function is chosen – its value is the BS utility at equilibrium.

3) *In the fully non-cooperative model, the players in equilibrium choose $\Psi_1 = \Psi_1^*$ and $\Psi_2 = \Psi_2^*$ satisfying*

$$c_{WC}^1(\Psi_1^*) = v = c_{WC}^2(\Psi_2^*) \qquad (11)$$

*and then both use a WC policy.*

What we see in this proposition is that when the BS can decide on the behavior of the users, it forces them to use 3G. In other cases (when users can decide on their behavior, but are given only partial information), the users' interest is to choose the CQI thresholds somewhere in the middle of the channel gain range. This can be seen as a desired trade-off between the global network performance at the equilibrium and the individual efficiency of all the users. On the other hand, the BS has an incentive to choose CQI thresholds either very high (first case in the Stackelberg scenario) or very low (the second case). Both these choices give little information for the user about actual channel condition, which is precisely what he wants to avoid. It is interesting and somewhat surprising that the optimal policy of the BS in the Stackelberg game can be both giving high or low values of CQI thresholds. This can however be explained when we understand the meaning of these two situations – very high value of the threshold means that no information about the channel state is given. In this case, when both users connect to 3G, this corresponds to the choice of the BS. Now, if in the "no information" case players choose WiFi, then the base station tries to divide the range of $h_i$ into a small (in terms of probability) part when the players use WiFi and, a large one when they use 3G. This is done by giving the lowest possible CQI threshold above which the players would have an incentive to use rather 3G than WiFi. This explains why the BS has an incentive to choose CQI thresholds very low in this case.

The final two results of this section are given without proofs, which are straightforward.

**Corollary 1.** *Note that the maximum network utility, obtained in scenario 1) is equal to $\beta_1 + \beta_2$. Obviously the utilities obtained in the other two scenarios always satisfy*

$$\beta_1 + \beta_2 \geq U_{BS}(\mathbf{P}^{BSE}, \Psi_i^{BSE}, \Psi_j^{BSE}) \qquad (12)$$
$$> U_{BS}(\mathbf{P}^{NC}, \Psi_i^{NC}, \Psi_j^{NC}) > 0$$

The important fact the corollary implies is that the users never choose the same way the base station would, but their interests however need not be necessarily contradicting. In the second corollary, we give the method to compute the price of anarchy(PoA) [15][16] for our model.

The PoA measures how good the system performance is when users play selfishly and reach the NE instead of playing to achieve the social optimum [15][16]. Note that as the maximum network utility which can be obtained is $\beta_1 + \beta_2$, the price of anarchy when players use strategy profile $\mathbf{P}$ is

$$PoA = \frac{\beta_1 + \beta_2}{U_{BS}(\mathbf{P}, \Psi_1, \Psi_2)}.$$

Thus, Proposition 3 implies that

**Corollary 2.** *The price of anarchy in the Stackelberg model equals 1 whenever $C_{CC}^i(\infty) > v$ for $i = 1, 2$. When for some $i$, $C_{CC}^i(\infty) \leq v$, then the price of anarchy is equal to the smaller of the two values:*

$$\min_{C_{WC}^k(1)(\Psi_1,\Psi_2) \geq v \geq C_{WC}^k(0)(\Psi_1,\Psi_2), k=1,2} \frac{\beta_1 + \beta_2}{\beta_1 e^{-\lambda_1 \Psi_1} + \beta_2 e^{-\lambda_2 \Psi_2}},$$

$$\min_{C_{WC}^i(0)(\Psi_1,\Psi_2) \geq v, C_{CC}^j(1)(\Psi_1,\Psi_2) \geq v} \frac{\beta_1 + \beta_2}{\beta_i + \beta_j e^{-\lambda_j \Psi_j}}.$$

*In fully non-cooperative model,*

$$PoA = \frac{\beta_1 + \beta_2}{\beta_1 e^{-\lambda_1 \Psi_1^*} + \beta_2 e^{-\lambda_2 \Psi_2^*}},$$

*where $\Psi_1^*$ and $\Psi_2^*$ satisfy (11).*

The above corollary is just a rewriting of the Proposition 3 using different language.

### IV. THE MULTI-USER CASE

Now let us consider the case where instead of two we have $n$ users choosing to connect either to WiFi or to 3G network. Again we assume that the information about the channel quality that user $i$ possesses is limited to that about the distributions of states $(s_j, b_i)$ of each of the players (including $i$), that is about $\alpha_j$ (or $\lambda_j$) and $\beta_j$ and to exact information about his own current state $(s_i, b_i)$ (but not about exact value of $h_i$). Our additional assumption about the model considered in this section is that the model is symmetric, that is all the values $\beta_i, \lambda_i$ and $\Psi_i$ defining it, are the same for each of the players (and equal to $\beta, \lambda$ and $\Psi$ respectively). This significantly simplifies the notation without any serious limitation of generality (we believe that some counterparts of all our results will be true also for asymmetric model).

To define the utilities of the players first let us redefine throughput for each system:

$$Thp_i^C = \log\left(1 + \frac{p h_i a_i b_i}{\sigma^2 + p \sum_{j \neq i} h_j a_j b_j}\right) \quad (13)$$

$$Thp^W = v \quad (14)$$

Again we assume that each of the players uses one of the three policies $WW, WC, CC$, where first letter stands for a player's action when his channel is bad, and the second one when his channel is good. As it is troublesome to write down the policies for each of $n$ players, we will make use of the fact that the game is symmetric, writing instead of the policy profile a policy statistics $\mathbf{K} = [k_{CC}, k_{WC}]$ with $k_{CC}$ denoting the number of players applying policy $CC$ and $K_{WC}$ – of players applying $WC$. Of course the number of those using policy $WW$ is $n - k_{CC} - k_{WC}$, so we will omit it. Given $\mathbf{K}$, we can define user $i$'s utility in state $s = 0, 1$ as[5]

$$u_i(s, \mathbf{K}) = \begin{cases} v; & \text{if user } i \text{ chooses } W \text{ at state } s, \\ C_{\mathbf{K}_{-i}}^i(s); & \text{if user } i \text{ chooses } C \text{ at state } s \end{cases} \quad (15)$$

[5] Notation $\mathbf{K}_{-i}$ used below denotes policy statistics defined as in the two-user case but without policy of user $i$.

where the functions $C_{\mathbf{K}_{-i}}^i$, describing the utility of player $i$ using 3G when his opponents use policies described by $\mathbf{K}$, are similarly as for the two-user case:

$$C_{\mathbf{K}_{-i}}^i(1) = \mathbb{E}[c_{\mathbf{K}_{-i}}^i(h_i) | h_i > \Psi] \quad (16)$$

$$= \frac{1}{\alpha_i} \int_{\Psi_i}^{\infty} c_{\mathbf{K}_{-i}}^i(h_i) \lambda_i e^{-\lambda_i h_i} dh_i,$$

$$C_{\mathbf{K}_{-i}}^i(0) = \mathbb{E}[c_{\mathbf{K}_{-i}}^i(h_i) | h_i < \Psi] \quad (17)$$

$$= \frac{1}{1 - \alpha_i} \int_0^{\Psi_i} c_{\mathbf{K}_{-i}}^i(h_i) \lambda_i e^{-\lambda_i h_i} dh_i.$$

Next, the functions $c_{\mathbf{K}_{-i}}^i$, defining utility of player $i$ using $C$ when channel gain is $h_i$ against policies $\mathbf{K}$ of his opponents, can be written as[6]:

$$c_{[k_1,k_2]}^i(h) = \sum_{r=0}^{k_1} \sum_{q=0}^{k_2} \sum_{v=0}^{q} \beta^{r+q}(1-\beta)^{k_1+k_2-r-q} \binom{k_1}{r}\binom{k_2}{q}$$

$$\binom{q}{v} \int_{[\Psi,\infty)^v \times \mathbb{R}^{n-q-1} \times [0,\Psi)^{q-v}} e^{-\lambda \sum_{j=1}^{n-1} h_j}$$

$$\log\left(1 + \frac{ph}{\sigma^2 + p \sum_{j=1}^{r+v} h_j}\right) \lambda^{n-1} dh_1 \ldots dh_{n-1}$$

Below, we give a generalization of Proposition 1 for the $n$-user case.

**Proposition 4.** *The symmetric $n$-user game considered in the paper always has a pure-strategy Bayes-Nash equilibrium of one of six types:*

(a) *When $C_{[0,0]}^i(1) \leq v$ then the profile where all the players use policy $WW$ is an equilibrium.*

(b) *When $C_{[0,k-1]}^i(1) \geq v \geq C_{[0,k]}^i(1)$ and $C_{[0,k-1]}^i(0) \leq v$, then any profile where $k$ players use policy $WC$ and all the others play $WW$ is an equilibrium.*

(c) *When $C_{[0,n-1]}^i(1) \geq v$ and $C_{[0,n-1]}^i(0) \leq v$ then the profile where all the players use policy $WC$ is an equilibrium.*

(d) *When $C_{[k,n-k-1]}^i(1) \geq v$ and $C_{[k-1,n-k]}^i(0) \geq v \geq C_{[k,n-k-1]}^i(0)$ then any profile where $k$ players apply policy $CC$ and the remaining $n - k$ players use policy $WC$ is an equilibrium.*

(e) *When $C_{[n-1,0]}^i(0) \geq v$ then the profile where all the players use policy $CC$ is an equilibrium.*

(f) *When $C_{[k-1,n-k]}^i(0) \geq v \geq C_{[k,n-k-1]}^i(1)$ then any profile where $k$ players apply policy $CC$ and the remaining $n - k$ players use policy $WW$ is an equilibrium.*

*It may also have another pure strategy Bayes-Nash equilibrium with $k$ players using $CC$ and $l < n - k$ using $WC$ when $C_{[k,l-1]}^i(1) \geq v \geq C_{[k,l]}^i(1)$ and $C_{[k-1,l]}^i(0) \geq v$.*

[6] Of course this formula is a generalization of the formulas for $c_k^i$ given in section III and it applies for any $n \geq 2$, in particular $c_{CC}^i \equiv c_{[1,0]}^i$, $c_{WC}^i \equiv c_{[0,1]}^i$ and $c_{WW}^i \equiv c_{[0,0]}^i$ when $n = 2$ and players are symmetric.

We give a corollary to this proposition. It gives a kind of consistency property for equilibria in games for different values of $n$.

**Corollary 3.**

*(a) Suppose that a profile where at least one player uses policy $WW$ and the number of players using policies $CC$ and $WC$ is $k$ is an equilibrium in $n$-user symmetric game. Then it is also an equilibrium in any $m$-user game defined with the same parameters $\beta, \lambda$ and $\Psi$ and $m \geq k$.*

*(b) Moreover for any fixed parameters $\beta, \lambda$ and $\Psi$ there exists an $n$ such that for any $m > n$ at least $m - n$ players use policy $WW$ in any equilibrium in $m$-user game.*

*Proof:* Note that $C^i_{[k_1,k_2]}(s)$ does not depend on the number of players in the game $n$, only on the number of those who use one of the policies $WC$ or $CC$. Just this implies part (a). Part (b) is due to the fact that $C^i_{[0,n-1]}(1) \to 0$ as $n \to \infty$. ∎

The next proposition generalizes the results for hierarchical model included in Proposition 3 for $n$-user symmetric games. We only consider scenarios 1) and 2) discussed there, as it is difficult to apply scenario 3) to the symmetric model. The BS utility, defined as before, as the expected number of players using 3G network, can be now written as:

$$U_{BS}([k_1, k_2], \Psi) = \beta(k_1 + e^{-\lambda \Psi} k_2).$$

**Proposition 5.**

1) *In the centralized model, the base station chooses any value of $\Psi$ and $CC$ policies for all the users.*
2) *In the Stackelberg model the base station computes for every $k \leq n$*

$$C^i_{[k-1,0]}(\infty) := \int_0^\infty c^i_{[k-1,0]}(h) \lambda e^{-\lambda h} dh$$

*and*

$$C^i_{[0,k-1]}(\infty) := \int_0^\infty c^i_{[0,k-1]}(h) \lambda e^{-\lambda h} dh,$$

*and finds $k^*$ and $n^*$ such that*

$$C^i_{[k^*-1,0]}(\infty) \geq v \geq C^i_{[k^*,0]}(\infty)$$

*and*

$$C^i_{[0,n^*-1]}(\infty) \geq v \geq C^i_{[0,n^*]}(\infty)$$

*Next:*

(a) *If $n \leq k^*$ then at the equilibrium the base station chooses any $\Psi$ such that $C^i_{[n-1,0]}(0)(\Psi) > v$ and all the players use policy $CC$.*

(b) *If $n > k^*$ then the base station computes for any $k$ such that $k^* \leq k \leq \min\{n, n^*\}$ and any $0 \leq l \leq k$, a $\Psi(k, l)$ such that*

$$\min\{C^i_{[l,k-l-1]}(1)(\Psi(k,l)), C^i_{[l-1,k-l]}(0)(\Psi(k.l))\} = v.$$

*If such a $\Psi(k,l)$ does not exist, or $v < C^i_{[l,k-l]}(1)$ and $k < n$, then it puts $P(k,l) = 0$. Otherwise it computes*

$$P(k,l) = \beta(l + e^{-\lambda \Psi(k,l)}(k-l)).$$

*Finally it chooses $k_{max}$ and $l_{max}$ with the biggest value of $P(k,l)$ (which equals the BS utility at equilibrium). The choice of $\Psi(k_{max}, l_{max})$ at the first stage and any profile of policies where $l_{max}$ players use policy $CC$ and $k_{max} - l_{max}$ play $WC$ will then be an equilibrium.*

We give one corollary to this proposition.

**Corollary 4.** *The price of anarchy in the $n$-user Stackelberg model can be computed as*

$$PoA = \frac{n\beta}{U_{BS}(\mathbf{K}, \Psi)}$$

*(where $n\beta$ is the maximum value of the base station's utility obtained in scenario 1) of Proposition 5), and is either equal to 1 when $k \leq k^*$, or satisfies*

$$PoA = \min_{k^* \leq k \leq n, k \leq n^*, 0 \leq l \leq k} \frac{n\beta}{P(k,l)}$$

*with $k^*$ and $n^*$ defined as in Proposition 5. For $n > n^*$ it grows to infinity linearly.*

The first part of this corollary is again just a rewriting of the results from Proposition 5 with the stress made on network utilities rather than strategies of the players. It shows that exactly the same procedure, used to find the equilibrium policies, can be applied to evaluate the performance of the network.

The second part of the corollary (unbounded increase of PoA for $n^*$) is a consequence of the fact that adding each new player to the game gives the BS more patterns of behavior of the users which can be stimulated by a proper choice of $\Psi$ only up to the threshold number of players $n^*$. From then on no new player in the game is interested in using 3G network, because for such a large number of players the throughput would be too degraded, regardless of how good the channel would be.

If someone is interested not in finding the equilibria for all the numbers of players, but only in the limit number of players for which the base station can lead the players to the desired equilibrium just by a proper choice of the CQI threshold, he may instead of computing $k^*$ given above an upper bound can be computed as given below.

**Corollary 5.** *The value $k^*$ appearing above can be bounded from above by $k^{**} = \max\{\frac{2}{\beta} + 1, k^+\}$ with $k^+$ satisfying*

$$\frac{\lambda}{2} e^{-\frac{(k^+-1)\beta^2}{2}} \int_0^\infty \log\left(1 + \frac{ph}{\sigma^2}\right) e^{-\lambda h} dh + \log\left(\frac{(k^+-1)\beta}{(k^+-1)\beta - 2}\right) = v.$$

This last bound cannot be given in a closed form, but it can be computed much faster than $k^*$.

## V. NUMERICAL ILLUSTRATIONS

We consider a simple scenario of an operator providing subscribers with a service available through a large 3G cell coexisting with a WiFi access point with constant throughput $v$. As mentioned before, users are characterized by the distribution of their 3G downlink channel and the distribution of their demand. We consider for each user a Rayleigh distributed channel fading. In order to validate our theoretical findings, we obtain users' actions at the equilibrium defined by users decisions to connect to WiFi or 3G at low/high channel state. In order to provide with extensive results we study the scenario of the multi-users case with increasing number of users. This scheme allows us to address the proposed distributed decision making problem and gain insights into how to design association policies in such a radio environment. Without loss of generality, we set $\lambda = 0.6$ as average channel state for all users and consider normalized user's CQI threshold $\Psi \in [0, 1]$. Unless otherwise stated, for all numerical applications, we assume the following numerical normalized values: $v = 0.25$ Mbits/sec, $\beta = 0.5$ and $\alpha$ derives from $\lambda$ and $\Psi$. It is then possible to compute the non-cooperative Bayes-Nash equilibrium strategies and the related users' utilities obtained at the equilibrium. For the hierarchical Stackelberg equilibrium, given the action of the BS, i.e., the CQI threshold $\Psi$, we compute the best-response function of the mobile users, i.e., the action of the mobile users which maximizes their utilities given the action of the 3G BS. The network utility is defined as the average throughput obtained by a user selecting the 3G BS. Finally, under the formerly defined policy statistics $\mathbf{K} = \{k, l\}$, the ratio number of user connected to system $\mathbf{S}$ (with $\mathbf{S} = C$ for the macro-cell and $\mathbf{S} = W$ for the WiFi AP), $\mathcal{L}(\mathbf{S})$, can be respectively expressed as follows: $\mathcal{L}(C) = (k + l\alpha)/n$ and $\mathcal{L}(W) = (n - k - l\alpha)/n$.

We first notice that, different equilibria can be archived based on the different scenarios as Proposition 5 points out. Secondly, we indeed observe that in the centralized case, no matter the number of interacting users, the base station always drive the users to select $CC$ for any value of $\Psi$. Meanwhile, as claimed in Prop. 5, we find that for low values of $n \leq k^*$, the hierarchical framework drives the users to select the $CC$ strategy when the channel quality threshold satisfies the inequality condition of the proposition. It means that when the channel is good enough to have better utility all users should select to attach with the base station for low number of users. As $\Psi$ increases, users tends to connect to 3G using policy $WC$. Asymptotically, when $\Psi$ grows large, users choose policy $CC$ at the equilibrium. This is illustrated in Figure 1 that depicts the load of both systems when $\Psi$ increases.

It can be observed that as $\Psi$ increases users choose to connect to 3G, increasing abruptly 3G load. This is due to the fact that for low values of $\Psi$, users may know that the channel is in a really bad state when the channel state is "bad" but do not know much about the quality of the channel when the state is "good". On the other way, when $\Psi$ is very high, "good" channel state means that the channel is in a very good state, and "bad" channel state means hardly any information about the channel quality. Accordingly, for low values of $\Psi$ users always choose WiFi in "bad" state and for high value of $\Psi$ always choose 3G in "good" state. This suggests that knowing the distribution of the channel (through $\alpha$), one can maximize the user throughput by correctly choosing the CQI threshold. The same observation is done for higher values of $n > k^*$ in figure 2, but here the threshold value of $\Psi$ is clearly identified by the conditions of Prop. 5. At this point, although higher values of $\Psi$ describe a better channel conditions, following this information is already misleading to users due to the large number of interacting users.

Interestingly, we have shown in this paper by means of a Stackelberg formulation that in order to make users connecting to 3G network, the BS could control the CQI thresholds rather than increasing the offered throughput. As a result, the operator not only obtains a better revenue at no additional cost but also significantly improves its energy efficiency.

To go further with the analysis, we resort to study the $PoA$ of the considered configurations of hierarchical and centralized schemes for numerically evaluated examples in order to give insight on the performance of the proposed hierarchical approach. From Figure 3, we may draw that the PoA is equal to 1 before $k^*$ as the hierarchically coordinated actions of 3G BS and users allow to identify not only the actual channel quality and traduces in good users' utility due to low number of users interacting. From $k^*$ to $n^*$ (defined in Prop. 5) as stated before, by increasing the number of users, each new user to the game gives the BS more patterns of behavior of the users which can be stimulated by a proper choice of $\Psi$ but only up to a number $n = n^*$. In our simulations, for $n > n^* = 41$ because of the large number of interacting users, the 3G BS throughput is so degraded that no new user will connect with the 3G regardless of how good is the channel. This is traduced in the figure 3 by a linear increase of the $PoA$.

These price of anarchy results offer hope that such a robust and accurate modeling can be designed around competition, because Stackelberg behavior does not arbitrarily degrade the mechanism's performance like the selfish does. However we have identified a threshold number of users above which no successful coordination can be achieved.

## VI. CONCLUSION

Motivated by the fact that in game theory it is well known that performance at equilibrium is not monotone increasing in the amount of information, we have proposed a hierarchical association method that combines benefits from both decentralized and centralized design in which the network operator dynamically chooses the offset about the state of the channel. The results of this study lead naturally to several lines of future investigation. The users' decision making is based on partial information that is signaled to the mobiles by the base station. A central design aspect is then for the base stations to decide how to aggregate information which then determines what to signal to the users. In this setting, we have shown that, in order to maximize its revenue, the network operator rather than increasing its offered throughput (which is costly) has an incentive to choose channel quality indicator thresholds either very low or very high. This may make the information given to the user when attempting to connect *misleading* since the throughput of a user cannot be directly inferred from the quality of his channel but also depends on the channel quality indicator thresholds (offsets) the base station broadcasts. In particular, there may be different equilibria (so different outcomes) depending on what information the base station broadcasts to users. We have finally analyzed the global performance indicators of the network. It shows that exactly the same procedure, used to find the equilibrium policies, can be applied to evaluate the performance of the network. Typically, we have characterized the price of anarchy on which we have derived an upper bound. It has been shown that the proposed approach provides a reasonable trade-off between centralized vs decentralized optimization in terms of the signaling overhead and the resulting network throughput performance.

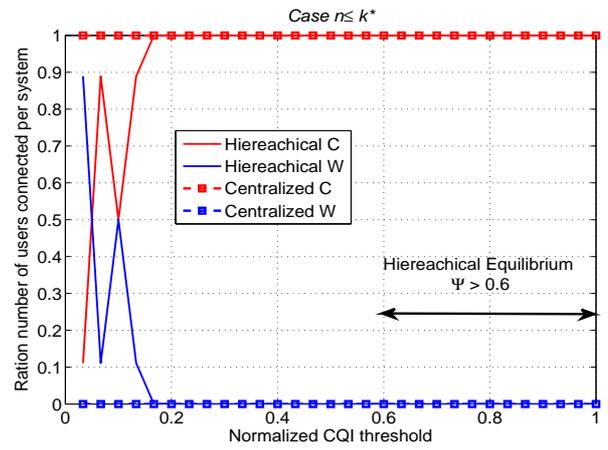

Fig. 1. Load of both WiFi and 3G systems as function of $\Psi$. Case $n = 9 \leq k* = 12$

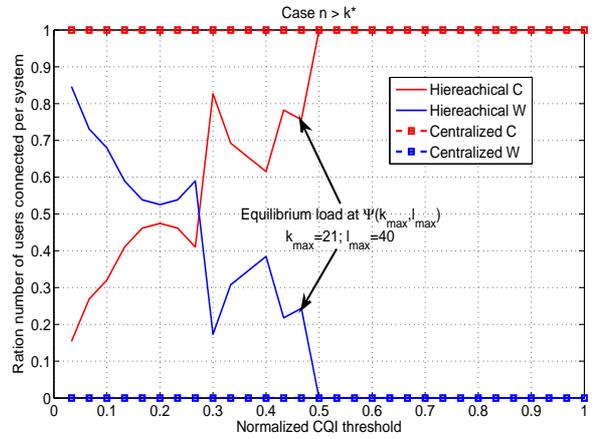

Fig. 2. Load of both WiFi and 3G systems as function of $\Psi$. Case $n = 40 > k* = 12$

## Appendix

*A. Proof of Proposition 3*

*Proof:*

1) is obvious and needs no explanation.
2) Since when $\Psi_1 \to \infty$ and $\Psi_2 \to \infty$, $C_{CC}^i(1)(\Psi_1, \Psi_2) \to C_{CC}^i(\infty)$ for $i = 1, 2$, when $C_{CC}^i(\infty) > v$, then for $\Psi_i$ large enough also $C_{CC}^i(0)(\Psi_1, \Psi_2) \geq v$ for $i = 1, 2$. But this means that $(CC, CC)$ is an equilibrium in the game at the second stage. Thus whenever $\Psi_1 < \Psi_1^{**}$ and $\Psi_2 < \Psi_2^{**}$ with $\Psi_i^{**}$ satisfying $C_{CC}^i(1)(\Psi_1^{**}, \Psi_2^*) = v$, the outcome of the Stackelberg game is that both players use 3G with probability 1, which gives the biggest value possible of the base station's utility. Now suppose that $C_{CC}^i(\infty) \leq v$. Then for any value of $\Psi_i$, playing $(CC,CC)$ is not an equilibrium in the game of the second stage. Thus, to maximize the expected number of players using 3G, the base station has to choose the $\Psi_i$ and $\Psi_j$ in such a way that the equilibrium in the game of the second stage was either $(WC,WC)$ or $(CC,WC)$ and the 3G BS utility function $U_{BS}$ (equal to $\beta_1 e^{-\lambda_1 \Psi_1} + \beta_2 e^{-\lambda_2 \Psi_2}$ in the first case and $\beta_i + \beta_j e^{-\lambda_j \Psi_j}$ in the second one) is the highest possible. This is done by solving the optimization problems defined in the proposition (the first problem for the case $(WC,WC)$, the second one for $(CC,WC)$).

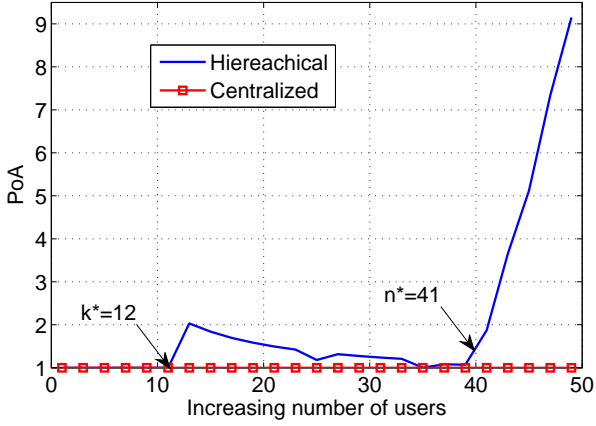

Fig. 3. Price of anarchy between hierarchical and centralized scheme.

3) First note that whenever $\Psi_1^*$ and $\Psi_2^*$ are chosen as in (11), $(WC,WC)$ is an equilibrium. This is because $C_{WC}^i(0)$ is the conditional expectation of $c_{WC}^i(h_i)$ over the set $H_- := \{c_{WC}^i(h_i) \leq v\}$, so it is definitely smaller than $v$. Similarly, $C_{WC}^i(1)$ is the conditional expectation of $c_{WC}^i(h_i)$ over the set $H_+ := \{c_{WC}^i(h_i) \geq v\}$, so it is bigger than $v$. Thus the condition for $(WC,WC)$ to be an equilibrium is definitely satisfied.

Now note that whenever player $i$ chooses $\Psi_i < \Psi_i^*$ at the first stage, but continues to use policy $WC$ in the second, he loses

$$\int_{\Psi_i}^{\Psi_i^*} (v - c_{WC}^i(h_i)) \lambda_i e^{-\lambda_i h_i} > 0.$$

Similarly, when he chooses $\Psi_i > \Psi_i^*$ he loses

$$\int_{\Psi_i^*}^{\Psi_i} (c_{WC}^i(h_i) - v) \lambda_i e^{-\lambda_i h_i} > 0.$$

On the other hand, when he changes both the $\Psi_i$ and the policy at the second stage, his utility is either $v$ (when he plays $WW$) or $\mathbb{E}[c_{WC}^i(h_i)]$ (when he uses policy $CC$), which are clearly both less than his current utility

$$P(h_i \in H_-) v + P(h_i \in H_+) \mathbb{E}[c_{WC}^i(h_i) | h_i \in H_+],$$

so $(\Psi_1^*, \Psi_2^*)$ is an equilibrium choice.

The last thing we need to show is that there are $\Psi_1^*$ and $\Psi_2^*$ satisfying (11). Let us construct functions[7] $\widehat{\Psi}_i(\Psi_j) := \{\Psi_i : c_{WC}^i(\Psi_i)(\Psi_j) = v\}$. It is immediate to see that since all the functions $c_k^i$ are nondecreasing, $\widehat{\Psi}_i$ are non-increasing functions from $[0, \infty)$ to itself. It is then obvious that the graphs of these two functions: $\{(\Psi_1, \Psi_2) : \Psi_1 = \widehat{\Psi}_1(\Psi_2)\}$ and $\{(\Psi_1, \Psi_2) : \Psi_2 = \widehat{\Psi}_2(\Psi_1)\}$ intersect, and thus (11) has a solution. ∎

*B. Proof of Proposition 4*

Before we prove Proposition 4, we need an auxiliary lemma.

**Lemma 1.** *For any $s = 0, 1$ the functions $C_{[k,l]}^i(s)(\Psi)$*

*(a) are decreasing in $k$, $l$,*

*(b) satisfy $C_{[k+1,l]}^i(s)(\Psi) \leq C_{[k,l+1]}^i(s)(\Psi)$,*

*(c) are increasing in $\Psi$.*

*Proof:* First note that

$$F(r,q) := \sum_{v=0}^{q} \binom{q}{v} \int_{[\Psi,\infty)^v \times \mathbb{R}^{n-q-1} \times [0,\Psi)^{q-v}} \lambda^{n-1}$$
$$\log\left(1 + \frac{ph}{\sigma^2 + p\sum_{j=1}^{r+v} h_j}\right) e^{-\lambda \sum_{j=1}^{n-1} h_j} dh_1 \ldots dh_{n-1}$$

is decreasing in $r$.

Next, we need to show that[8]

$$F(r+1, q) \leq F(r, q+1) \leq F(r, q) \text{ for any } r, q \geq 0 \quad (18)$$

---

[7] Here we use a convention that in the second bracket we give the value of player $j$'s threshold, which appears in the definitions of $c_k^i$, but was omitted so far.

[8] The first inequality will be used to prove part (b) of the lemma. The second one is the monotonicity of $F$ in $q$.

$$F(r,q+1) = \sum_{v=0}^{q+1} \binom{q+1}{v} \int_{[\Psi,\infty)^v \times \mathbb{R}^{n-q-2} \times [0,\Psi)^{q-v+1}} \lambda^{n-1} \cdot$$

$$\log\left(1 + \frac{ph}{\sigma^2 + p\sum_{j=1}^{r+v} h_j}\right) e^{-\lambda \sum_{j=1}^{n-1} h_j} dh_1 \ldots dh_{n-1}$$

$$= \sum_{v=0}^{q} \binom{q}{v} \int_{[\Psi,\infty)^v \times \mathbb{R}^{n-q-2} \times [0,\Psi)^{q-v}} \lambda^{n-2} \cdot$$

$$\left[ \int_0^\Psi \log\left(1 + \frac{ph}{\sigma^2 + p\sum_{j=1}^{r+v} h_j}\right) \lambda e^{-\lambda h_{r+v+1}} dh_{r+v+1} \right.$$

$$\left. + \int_\Psi^\infty \log\left(1 + \frac{ph}{\sigma^2 + p\sum_{j=1}^{r+v+1} h_j}\right) \lambda e^{-\lambda h_{r+v+1}} dh_{r+v+1} \right]$$

$$e^{-\lambda \sum_{j=1}^{n-2} h_j} dh_1 \ldots dh_{r+v} dh_{r+v+2} \ldots dh_{n-1}$$

$$\geq \sum_{v=0}^{q} \binom{q}{v} \int_{[\Psi,\infty)^v \times \mathbb{R}^{n-q-2} \times [0,\Psi)^{q-v}} \lambda^{n-2}$$

$$\int_0^\infty \log\left(1 + \frac{ph}{\sigma^2 + p\sum_{j=1}^{r+v+1} h_j}\right) \lambda e^{-\lambda h_{r+v+1}} dh_{r+v+1}$$

$$e^{-\lambda \sum_{j=1}^{n-2} h_j} dh_1 \ldots dh_{r+v} dh_{r+v+2} \ldots dh_{n-1}$$

$$= \sum_{v=0}^{q} \binom{q}{v} \int_{[\Psi,\infty)^v \times \mathbb{R}^{n-q-1} \times [0,\Psi)^{q-v}} \lambda^{n-1} \cdot$$

$$\log\left(1 + \frac{ph}{\sigma^2 + p\sum_{j=1}^{r+v+1} h_j}\right) e^{-\lambda \sum_{j=1}^{n-1} h_j} dh_1 \ldots dh_{n-1}$$

$$= F(r+1, q)$$

The second inequality in (18) is proved analogously, only the inequality

$$\left[ \int_0^\Psi \log\left(1 + \frac{ph}{\sigma^2 + p\sum_{j=1}^{r+v} h_j}\right) \lambda e^{-\lambda h_{r+v+1}} dh_{r+v+1} \right.$$
$$\left. + \int_\Psi^\infty \log\left(1 + \frac{ph}{\sigma^2 + p\sum_{j=1}^{r+v+1} h_j}\right) \lambda e^{-\lambda h_{r+v+1}} dh_{r+v+1} \right]$$
$$\leq \int_0^\infty \log\left(1 + \frac{ph}{\sigma^2 + p\sum_{j=1}^{r+v} h_j}\right) \lambda e^{-\lambda h_{r+v+1}} dh_{r+v+1}$$

is used instead of the one used above.

Now, to prove (a) of the lemma note that $c^i_{[k,l]}(h)$ is the expected value of $\sum_{q=0}^{l} \beta^q (1-\beta)^{l-q} \binom{l}{q} F(r,q)$ when $r$ is a random value with the binomial distribution $\text{Bin}(k, \beta)$. As distribution $\text{Bin}(k+1, \beta)$ strictly stochastically dominates $\text{Bin}(k, \beta)$, the expected value with respect to $\text{Bin}(k+1, \beta)$ of any decreasing function is smaller than that with respect to $\text{Bin}(k, \beta)$ and thus

$$c^i_{[k+1,l]}(h) < c^i_{[k,l]}(h)$$

for any $h \geq 0$. But, as $C^i_{[k,l]}(0)$ and $C^i_{[k,l]}(1)$ are conditional expectations of $c^i_{[k,l]}(h)$ over some fixed sets, this immediately implies that they are both decreasing in $k$. The fact that they are decreasing in $l$ is proved analogously – the only difference is that the monotonicity of $F$ in $q$ (instead of the monotonicity in $r$) is used.

To prove (b) first note that for any $k, l \geq 0$

$$c^i_{[k,l]}(h) = \sum_{p=0}^{k+l} \beta^p (1-\beta)^{k+l-p} \binom{k+l}{p} G(p, k, l),$$

where $G(p, k, l) = \sum_{a=\max\{0,p-l\}}^{\min\{p,k\}} \frac{\binom{k}{a}\binom{l}{p-a}}{\binom{k+l}{p}} F(a, p-a)$. Note however that $G(p, k, l)$ is the expected value of $F(a, p-a)$ when $a$ is a random variable with the hypergeometric distribution $\text{Hypergeometric}(k+l, k, p)$. Since $\text{Hypergeometric}(k+l+1, k+1, p)$ strictly stochastically dominates $\text{Hypergeometric}(k+l+1, k, p)$, and $F(a, p-a)$ is by (18) a decreasing function of $a$ for any fixed $p$, $G(p, k+1, l)$, which is the expected value of $F(a, p-a)$ with respect to that first distribution is not bigger than $G(p, k, l+1)$, which is the expected value of $F(a, p-a)$ with respect to the second one. But this immediately implies that also

$$c^i_{[k+1,l]}(h) = \mathbb{E}^{p \sim \text{Bin}(k+l+1, \beta)}[G(p, k+1, l)]$$
$$\leq \mathbb{E}^{p \sim \text{Bin}(k+l+1, \beta)}[G(p, k, l+1)] = c^i_{[k,l+1]}(h).$$

The same arguments as in part (a) imply that this inequality is preserved by $C^i_{[k+1,l]}(s)(\Psi)$ and $C^i_{[k,l+1]}(s)(\Psi)$.

To prove the last part of the lemma take $\Psi_1 < \Psi_2$ and define for any $q$ and $\alpha \in \{0, 1, 2\}^q$

$$S_\alpha(\Psi_1, \Psi_2) = \{(h_1, \ldots, h_{n-1}) \in \mathbb{R}^{n-1} :$$
$$0 \leq h_j < \Psi_1 \text{ if } \alpha_j = 0, \Psi_1 \leq h_j < \Psi_2 \text{ if } \alpha_j = 1,$$
$$\Psi_2 \leq h_j \text{ if } \alpha_j = 2\}.$$

Note that $\mathbb{R}^{n-1} = \bigcup_\alpha S_\alpha(\Psi_1, \Psi_2)$. Next note that $F(r, q)(\Psi_1)$ is the integral over $\mathbb{R}^{n-1}$ of the function $f_1$ defined on each $S_\alpha(\Psi_1, \Psi_2)$ separately, as

$$\log\left(1 + \frac{ph}{\sigma^2 + p\sum_{j \geq 1} h_j}\right) \lambda^{n-1} e^{-\lambda \sum_{j=1}^{n-1} h_j}.$$

On the other hand $F(r, q)(\Psi_2)$ is the integral over $\mathbb{R}^{n-1}$ of the function $f_2$ defined on each $S_\alpha(\Psi_1, \Psi_2)$ separately, as

$$\log\left(1 + \frac{ph}{\sigma^2 + p\sum_{j \geq 2} h_j}\right) \lambda^{n-1} e^{-\lambda \sum_{j=1}^{n-1} h_j}.$$

Clearly $f_1 < f_2$, and so $F(r, q)(\Psi_1) < F(r, q)(\Psi_2)$ for any $r$ and $q$. This immediately implies that also $c^i_{[k,l]}(h)(\Psi_1) < c^i_{[k,l]}(h)(\Psi_2)$. However, note that since $c^i_{[k,l]}(h)$ are also increasing in $h$: Similarly

$$C^i_{[k,l]}(0)(\Psi_1) = \mathbb{E}[c^i_{[k,l]}(h)(\Psi_1) | h < \Psi_1]$$
$$\leq \mathbb{E}[c^i_{[k,l]}(h)(\Psi_1) | h < \Psi_2]$$
$$< \mathbb{E}[c^i_{[k,l]}(h)(\Psi_2) | h < \Psi_2]$$
$$= C^i_{[k,l]}(0)(\Psi_2),$$

which ends the proof of lemma. ∎

Now we are able to prove Proposition 4.

*Proof:* First note that it is clear from the definition of $c^i_{[k,l]}(h)$ that this is an increasing function of $h$ and thus

$$C^i_{[k,l]}(0) < C^i_{[k,l]}(1) \qquad (19)$$

for any values of $k$ and $l$. Next it is enough to check the definition of Bayes-Nash equilibrium (inferring (19) and Lemma 1 if needed) that the sets of inequalities appearing in the proposition define respective equilibria. What is left to show is that cases (a–f) cover all the possible situations. Suppose that case (e) does not hold. Then, either

$$C^i_{[0,0]}(0) < v \qquad (20)$$

or there exists a $k$ such that

$$C^i_{[k-1,n-k]}(0) \geq v \geq C^i_{[k,n-k-1]}(0). \qquad (21)$$

Clearly (20) is covered by cases (a–c) of Proposition 4. On the other hand (21) implies either case (d) of Proposition 4 or the following inequality (here (19) is used):

$$C^i_{[k-1,n-k]}(0) \geq v > C^i_{[k,n-k-1]}(1),$$

which is exactly the case (f) of the proposition. ∎

### C. Proof of Proposition 5

*Proof:*
Part 1) is obvious. 2) Since when $\Psi \to \infty$, $C^i_{[n-1,0]}(0)(\Psi) \to C^i_{[n-1,0]}(\infty)$, then if $C^i_{[n-1,0]}(\infty) > v$, for $\Psi$ large enough also $C^i_{[n-1,0]}(0)(\Psi) > v$, which means that all the players apply policy $CC$ in equilibrium at the second stage of the game. Thus whenever $\Psi$ is big enough, the outcome of the Stackelberg game is that all the players use 3G with probability 1, which gives the biggest value possible of the base station's utility.

Now suppose that $C^i_{[n-1,0]}(\infty) \leq v$. Then for any value of $\Psi$, not every player uses policy $CC$ at the equilibrium of the game of the second stage. Thus, to maximize the $U_{BS}$, the base station has to choose the $\Psi$ in such a way that at the equilibrium of the game of the second stage some (say $l$) players would apply policy $CC$ and some other (say $k-l$) would apply $WC$, and that the base station's utility was the highest possible. This is done by solving the optimization problems of finding the smallest $\Psi$ such that the profile $[k-l, l]$ is an equilibrium in the game defined by this $\Psi$, that is satisfying

$$C^i_{[l,k-l-1]}(1)(\Psi) \geq v \geq C^i_{[l,k-l]}(1)(\Psi) \quad \text{and} \quad C^i_{[l-1,k-l]}(0)(\Psi) \geq v$$

However, as by Lemma 1 $C^i_{\mathbf{K}}(s)(\Psi)$ are increasing functions of $\Psi$ for any fixed $\mathbf{K}$, this maximum is achieved for $\Psi$ satisfying (2b). When the values of $U_{BS}$ for each such $\Psi$ are computed, and the biggest one of them is chosen, this is certainly the biggest value of the base station's utility that can be obtained in the Stackelberg scenario. ∎

### D. Proof of Corollary 5

*Proof:*
Let us assume that

$$k > \frac{2}{\beta} + 1. \qquad (22)$$

First note that $C^i_{[k-1,0]}(\infty)$ is

$$\sum_{r=0}^{k-1} \beta^r (1-\beta)^{k-1-r} \binom{k-1}{r} \mathbb{E}[\log\left(1 + \frac{ph}{\sigma^2 + p\sum_{j=1}^r h_j}\right)] \qquad (23)$$

where $h$ and $h_1, \ldots, h_{k-1}$ are independent exponentially distributed random values with common parameter $\lambda$.

Next let $r^* = \frac{(k-1)\beta}{2}$. Now (23) can be rewritten as

$$\sum_{r<r^*} \beta^r (1-\beta)^{k-1-r} \binom{k-1}{r} \mathbb{E}[\log\left(1 + \frac{ph}{\sigma^2 + p\sum_{j=1}^r h_j}\right)]$$
$$+ \sum_{r \geq r^*} \beta^r (1-\beta)^{k-1-r} \binom{k-1}{r} \mathbb{E}[\log\left(1 + \frac{ph}{\sigma^2 + p\sum_{j=1}^r h_j}\right)].$$

Since the function $\mathbb{E}[\log\left(1 + \frac{ph}{\sigma^2 + p\sum_{j=1}^r h_j}\right)]$ is clearly positive decreasing, the first element of this sum can be bounded from above by

$$\text{Prob}[r < r^*] E[\log\left(1 + \frac{ph}{\sigma^2}\right)],$$

where $\text{Prob}[r < r^*]$ is the probability that a random value with binomial distribution $Bin(k-1, \beta)$ is smaller than $r^*$. This probability, using Hoeffding's inequality [17] can be bounded above by $\frac{1}{2}e^{-\frac{(k-1)\beta^2}{2}}$ and thus the whole term by $\frac{1}{2}e^{-\frac{(k-1)\beta^2}{2}} E[\log\left(1 + \frac{ph}{\sigma^2}\right)]$.

Analogously, the second element of the sum can be bounded from above by

$$\text{Prob}[r \geq r^*] \mathbb{E}[\log\left(1 + \frac{ph}{\sigma^2 + p\sum_{j=1}^{r^*} h_j}\right)]$$

and further by

$$\mathbb{E}[\log\left(1 + \frac{h}{\sum_{j=1}^{r^*} h_j}\right)]. \qquad (24)$$

Now note that $1 + \frac{h}{\sum_{j=1}^{r^*} h_j}$ is a random value with Pareto distribution [18, Chap. 20, Sec. 12] with parameters 1 and $r^*$, whose average is (for $r^* > 1$, which is guaranteed by our assumption (22)) $\frac{r^*}{r^*-1}$. Since logarithm is a concave function, we can use Jensen's inequality to bound (24) from above by

$$\log\left(\frac{r^*}{r^*-1}\right) = \log\left(\frac{(k-1)\beta}{(k-1)\beta-2}\right).$$

This implies that

$$C^i_{[k-1]}(\infty) < \frac{\lambda}{2} e^{-\frac{(k-1)\beta^2}{2}} \int_0^\infty \log\left(1 + \frac{ph}{\sigma^2}\right) e^{-\lambda h} dh$$
$$+ \log\left(\frac{(k-1)\beta}{(k-1)\beta-2}\right)$$

and consequently that for any $k$ such that the RHS of the above inequality equals $v$ the LHS will be smaller than $v$ and thus $k > k^*$. ∎